\documentclass[12pt]{article}
\usepackage{bm}
\usepackage{amsmath}
\usepackage{amssymb}
\usepackage{float}
\usepackage{url}

\newcommand{\nn}{\nonumber\\}

\renewcommand{\thepage}{}
\makeatletter
\@addtoreset{equation}{section}
\renewcommand{\theequation}{\thesection.\@arabic\c@equation}
\makeatother
\renewcommand{\thefootnote}{\fnsymbol{footnote}}
\begin{document}
\begin{titlepage}
\title{
\vspace*{-4ex}
\hfill
\begin{minipage}{3.5cm}
\end{minipage}\\
\bf Open String Fields as Matrices
\vspace{0.5em}
}

\author{
Isao {\sc Kishimoto},$^{1}$
Toru {\sc Masuda},$^{2}$
Tomohiko {\sc Takahashi},$^{2}$\\
and
Shoko {\sc Takemoto}$^{2}$
\\
\vspace{0.5ex}\\
$^1${\it Faculty of Education, Niigata University,
Niigata 950-2181, Japan}\\
$^2$
{\it Department of Physics, Nara Women's University,}
{\it Nara 630-8506, Japan}}
\date{}
%
\maketitle
%

\begin{abstract}
\normalsize
We show that the action expanded around Erler-Maccaferri's $N$ D-brane
 solution describes the $N+1$ D-brane system where one D-brane
 disappears due to tachyon condensation. 
String fields on multi-branes can be regarded as block
 matrices of a string field on a single D-brane in the same way as
matrix theories. 
\end{abstract}
\end{titlepage}

\renewcommand{\thepage}{\arabic{page}}
\renewcommand{\thefootnote}{\arabic{footnote}}
\setcounter{page}{1}
\setcounter{footnote}{0}
%
\section{Introduction}

Open string field theory has the possibility of revealing non-perturbative
aspects of string theory. Recently, Erler and Maccaferri have proposed a
method to construct classical solutions, which
are expected to describe any open string background \cite{Erler:2014eqa}.
This indeed implies that open string field theory is able to 
give a unified description of various D-branes regarded
as non-perturbative objects of string theory.

Multi-brane solutions in Ref.~\cite{Erler:2014eqa}
provide a correct vacuum energy and gauge invariant
observables. Accordingly, in order
to prove whether the theory describes a multi-brane system, it is
necessary to clarify open and closed string spectra 
in the background of the solution. However, it is difficult to give
a definite answer to this problem, because there are some
subtleties concerning BRST cohomology in
the background \cite{Erler:2014eqa}.

There is another question related to the degree of freedom of string
fields in the background.
We have one string field in the theory on a single D-brane.
However, in the case that
the multi-brane solution provides the background of
the $N$ D-branes, the number of string fields increases to  $N^2$
around the solution. Intuitively, $N^2$ fluctuation fields in the
multi-brane background seem to be introduced as redundant degrees of
freedom. Here, it is natural to ask how to generate $N^2$
string fields or Chan-Paton factors from one string field.

On the other hand, it is well-known that matrix theories are able to
describe various D-branes \cite{Banks:1996vh,Ishibashi:1996xs}.
In matrix theories, D-branes are created by classical solutions as block
diagonal matrices. After expanding a matrix around the solution, block
matrices can be understood as representing open strings connecting each
D-brane. Here, it should be noted that there are similarities between
the matrix and the open string field: the matrix is deeply tied to the 
open string degree of freedom and an open string field is interpreted as
a matrix in which the left and right indices correspond to the left and
right half-strings \cite{Witten:1985cc}. Then, it seems plausible that
$N^2$ string fields on $N$ D-branes are embedded like block matrices
in a string field on a D-brane.

The purpose of this paper is to clarify the origin of the $N^2$ string
fields in the background of an $N$ D-brane solution.
We will show that the theory expanded around the solution is regarded as
an open string field theory on $N+1$
D-branes, but in which a D-brane vanishes as a result of tachyon
condensation. Then, the $N^2$ string fields will be given as block
matrices in a string field as an infinite-dimensional matrix.
Consequently, we can expect that the $N$ D-brane solution
correctly reproduces the open and closed string spectra in the $N$
D-brane background. 

The paper is organized as follows. In Sect.~\ref{sec2}, after
a brief explanation of multi-brane solutions by Erler-Maccaferri
\cite{Erler:2014eqa}, we will introduce projection operators acting on
a space of string fields. Then, we will analyze
a string field theory expanded around the $N$ D-brane  solution
in terms of the projectors. In Sect.~\ref{sec3}, we will give concluding
remarks.

\section{Open string field theory around multi-brane solutions
\label{sec2}}

\subsection{Erler-Maccaferri's solution for $N$ D-branes}

The action of bosonic cubic open string field theory is
\begin{eqnarray}
 S[\Psi;Q_{\rm B}]=-\frac{1}{g^2}\int
\left(\frac{1}{2}\Psi Q_{\rm B}\Psi+\frac{1}{3}
\Psi^3\right).
\label{action}
\end{eqnarray}
From the action, the equation of motion is given by
\begin{eqnarray}
 Q_{\rm B}\Psi+\Psi^2=0.
\label{eqm}
\end{eqnarray}

To construct multi-brane solutions for $N$ D-branes, Erler and
Maccaferri introduced $N$ pairs of
regularized boundary-conditions-changing operators, $\Sigma_a$ and
$\bar{\Sigma}_a$ $(a=1,\cdots, N)$
\cite{Erler:2014eqa}.\footnote{$\Sigma_a$ and $\bar{\Sigma}_a$ are
constructed by boundary-condition-changing (bcc) operators,
$\sigma_a$ and $\bar{\sigma}_a$, satisfying the operator product
expantion:
$\bar{\sigma}_a(z')\sigma_b(z)\to \delta_{ab}$ $(z'\to z)$. In the
Minkowski background, a zero momentum condition for the bcc operators
is not necessarily required. So, the simplest bcc operators are given
as
\begin{eqnarray}
 \sigma_a(z)=e^{ik_a\cdot X(z)},\ \ \ 
 \bar{\sigma}_a(z)=e^{-ik_a\cdot X(z)},
\end{eqnarray}
where $k^\mu_a$ satisfy $k_a^2=0$ and $k_a\cdot k_b<0\
(a\neq b)$.
 For example, we can take
$k_a^\mu=(a,1,\sqrt{a^2-1},0,\cdots,0)$.
}
These operators satisfy
\begin{eqnarray}
 \bar{\Sigma}_a\Sigma_b=\delta_{ab},
\label{SigmaBarSigma}
\end{eqnarray}
and
\begin{eqnarray}
 Q_{\rm T}\Sigma_a=Q_{\rm T}\bar{\Sigma}_a=0,
\label{QSigma}
\end{eqnarray}
where $Q_{\rm T}$ is a modified BRST operator on the tachyon vacuum.
From (\ref{QSigma}), we find that
\begin{eqnarray}
 Q_{\rm B}\Sigma_a=\Sigma_a\psi_{\rm T}-\psi_{\rm T}\Sigma_a,
\ \ \ 
 Q_{\rm B}\bar{\Sigma}_a=\bar{\Sigma}_a\psi_{\rm T}-\psi_{\rm T}
\bar{\Sigma}_a,
\label{QBSigma}
\end{eqnarray}
where $\psi_{\rm T}$ denotes the tachyon vacuum solution of
(\ref{eqm}). Here, we only assume that $\Sigma_a$ and
$\bar{\Sigma}_a$ satisfy Eqs.~(\ref{SigmaBarSigma})
and (\ref{QSigma}) (or equivalently (\ref{QBSigma}))
for a tachyon vacuum solution $\psi_{\rm T}$, 
regardless of wedge-based\cite{Schnabl:2005gv,Erler:2009uj} or
identity-based\cite{Takahashi:2002ez,Kishimoto:2002xi} solutions.

Using $\psi_{\rm T}$, $\Sigma_a$ and $\bar{\Sigma}_a$,
Erler-Maccaferri provided a multi-brane solution as\cite{Erler:2014eqa}
\begin{eqnarray}
 \Psi_0=\psi_{\rm T}
-\sum_{a=1}^N\Sigma_a\psi_{\rm T}\bar{\Sigma}_a.
\label{EMsol}
\end{eqnarray}
We can calculate the action for $\Psi_0$ with the help of 
(\ref{SigmaBarSigma}) and (\ref{QSigma}):
\begin{eqnarray}
 S[\Psi_0;Q_{\rm B}]=-(N-1)S[\psi_{\rm T};Q_{\rm B}].
\label{N-1 S}
\end{eqnarray}
Then, the solution $\Psi_0$ provides a
correct vacuum energy for $N$ D-branes.
Expanding the string field around the solution as $\Psi=\Psi_0+\psi$,
we can obtain the action for the fluctuation $\psi$:
\begin{eqnarray}
 S[\Psi;Q_{\rm B}]=S[\Psi_0;Q_{\rm B}]+S[\psi;Q_{\Psi_0}],
\end{eqnarray}
where the operator $Q_{\Psi_0}$ denotes the shifted BRST operator by the
solution $\Psi_0$.

\subsection{Projectors}

To clarify the physical interpretation of $S[\psi;Q_{\Psi_0}]$, we
introduce $N$ projection states as follows:
\begin{eqnarray}
 P_a=\Sigma_a\bar{\Sigma}_a\ \ \ (a=1,\cdots,N),
\end{eqnarray}
where the same indices $a$ are not summed.
Here we have to notice that, as pointed out in Ref.~\cite{Erler:2014eqa},
$\bar{\Sigma}_a$ should be multiplied to $\Sigma_a$ from the left
and so these projectors should be dealt with carefully. More precisely,
we define the projections for arbitrary string fields $A$ and $B$ as
follows:
\begin{eqnarray}
 AP_a B = (A \Sigma_a)(\bar{\Sigma}_a B).
\label{defP}
\end{eqnarray}
From (\ref{SigmaBarSigma}) and (\ref{defP}), we can easily find that
\begin{eqnarray}
 P_aP_bA=\Sigma_a(\bar{\Sigma}_a P_bA)=\Sigma_a((\bar{\Sigma}_a\Sigma_b)
(\bar{\Sigma}_bA))=\delta_{ab}P_aA.
\end{eqnarray}
This is a sufficient 
definition of the projectors for later calculation. But
it suggests that we need to insert some infinitesimal worldsheet to
separate $\Sigma_a$ and $\bar{\Sigma}_a$.
We will discuss this point further in the last section.

In addition to $P_a$, we define the $0$th projection as a complementary
projector:
\begin{eqnarray}
 P_0=1-\sum_{a=1}^{N}P_a,
\end{eqnarray}
where $1$ denotes the identity string field.
By definition, these $N+1$ projections satisfy
\begin{eqnarray}
 \sum_{\alpha=0}^NP_\alpha=1,
\end{eqnarray}
where the Greek indices are used for values $0,1,\cdots,N$.
From (\ref{QSigma}), it follows that $Q_{\rm T}P_\alpha=0$
and then we have
\begin{eqnarray}
Q_{\rm B}P_\alpha=P_\alpha\psi_{\rm T}-\psi_{\rm T}P_\alpha.
\label{QBPa}
\end{eqnarray}
Moreover, we can find some relations among $P_\alpha$, $\Sigma_\alpha$
and $\bar{\Sigma}_\alpha$:
\begin{eqnarray}
&&
 P_a\Sigma_b= \Sigma_a\delta_{ab},
\ \ \ 
\bar{\Sigma}_aP_b=\bar{\Sigma}_a\delta_{ab},
\ \ \ 
 P_0\Sigma_b= 0,
\ \ \ 
\bar{\Sigma}_aP_0=0.
\label{PaSigma}
\end{eqnarray}

With the help of these projectors, the string field $\Psi$ can be
partitioned into $(N+1)\times (N+1)$ blocks:
\begin{eqnarray}
 \Psi=\sum_{\alpha=0}^N\sum_{\beta=0}^NP_\alpha \Psi P_\beta=
\left(
\begin{array}{cccc}
 \Psi_{00}& \Psi_{01}& \cdots& \Psi_{0 N}\\ 
 \Psi_{10}& \Psi_{11}& \cdots& \Psi_{1 N}\\
 \vdots& \vdots& \ddots& \vdots\\
 \Psi_{N0}& \Psi_{N1}& \cdots& \Psi_{N N}
\end{array}\right),
\end{eqnarray}
where $\Psi_{\alpha\beta}$ is defined as the $(\alpha,\beta)$  sector
of $\Psi$, i.e., $\Psi_{\alpha\beta}\equiv P_\alpha\Psi P_\beta$.

According to Ref.~\cite{Erler:2014eqa}, the second term in (\ref{EMsol})
is a solution to the equation of motion at the tachyon vacuum.
From (\ref{PaSigma}), the second term is represented as
\begin{eqnarray}
-\sum_{a=1}^N\Sigma_a\psi_{\rm T}\bar{\Sigma}_a
=
\left(
\begin{array}{ccccc}
  0&  0& 0& \cdots& 0\\ 
  0& -\Sigma_1\psi_{\rm T}\bar{\Sigma}_1& 
0
& \cdots& 0\\
  0& 0& 
 -\Sigma_2\psi_{\rm T}\bar{\Sigma}_2& \cdots& 0\\
  \vdots &\vdots& \vdots& \ddots& \vdots\\
  0& 0& 0&  \cdots& -\Sigma_N\psi_{\rm
 T}\bar{\Sigma}_N\\
\end{array}\right).
\end{eqnarray}
Accordingly, it turns out that
the $N$ D-brane solution at the tachyon vacuum is given as a
block diagonal matrix. This is a similar result to
the case of matrix theories \cite{Banks:1996vh,Ishibashi:1996xs}. 

\subsection{Background described by the solution}

Now, we consider the fluctuation $\psi$ around the $N$ D-brane solution.
According to the previous subsection, $\psi$ can be written by matrix
representation:
\begin{eqnarray}
 \psi&=&\sum_{\alpha=0}^N\sum_{\beta=0}^N \tilde{\phi}_{\alpha\beta},
\label{psimatrix}
\end{eqnarray}
where $\tilde{\phi}_{\alpha\beta}=P_\alpha\psi P_\beta$.
$\tilde{\phi}_{\alpha\beta}$ represents a block matrix
of $\psi$ with infinite dimension.

Here, we consider change of variables of $\tilde{\phi}_{\alpha\beta}$.
$\tilde{\phi}_{ab}$ can be rewritten as
\begin{eqnarray}
 \tilde{\phi}_{ab}=P_a\tilde{\phi}_{ab}P_b=\Sigma_a(\bar{\Sigma}_a
\tilde{\phi}_{ab}\Sigma_b)\bar{\Sigma}_b.
\end{eqnarray}
So, we can change the variables from $\tilde{\phi}_{ab}$ to
$\phi_{ab}=\bar{\Sigma}_a\tilde{\phi}_{ab}\Sigma_b$.
Similarly, writing $\tilde{\phi}_{0a}=\chi_a\bar{\Sigma}_a$,
$\tilde{\phi}_{a0}=\Sigma_a\bar{\chi}_a$,
the fluctuation $\psi$ is represented as
\begin{eqnarray}
 \psi&=&
 \chi+ \sum_{a=1}^N\chi_a\bar{\Sigma}_a
+ \sum_{a=1}^N\Sigma_a\bar{\chi}_a
+\sum_{a=1}^N\sum_{b=1}^N\Sigma_a\phi_{ab}\bar{\Sigma}_b
\nn
&=&
\left(
\begin{array}{cc}
 \chi& \chi_b\bar{\Sigma}_b\\
&\\
 \Sigma_a\bar{\chi}_a &
 \Sigma_a\phi_{ab}\bar{\Sigma}_b
\end{array}\right),
\label{PsiTildePhi}
\end{eqnarray}
where we rewrite $\tilde{\phi}_{00}$ as $\chi$.

Similar to the equation  $Q_{\Psi_0}(\Sigma_a A\bar{\Sigma}_b)=
\Sigma_a(Q_{\rm 
B}A)\bar{\Sigma}_b$ given in Ref.~\cite{Erler:2014eqa}, by using
(\ref{QBSigma}) and (\ref{QBPa}), we have
\begin{eqnarray}
Q_{\Psi_0}(P_0AP_0)&=&P_0(Q_{\rm T}A)P_0,
\\
Q_{\Psi_0}(P_0A\bar{\Sigma}_a)&=&P_0(Q_{{\rm T}0}A)\bar{\Sigma}_a,
\\
Q_{\Psi_0}(\Sigma_aAP_0)&=&\Sigma_a(Q_{0{\rm T}}A)P_0,
\end{eqnarray}
where the operator $Q_{\psi_1\psi_2}$ is defined as
$Q_{\psi_1\psi_2}A=Q_{\rm B}A+\psi_1A-(-1)^{|A|}A\psi_2$ for two
classical solutions $\psi_1$ and $\psi_2$ \cite{Erler:2014eqa}, and then
$Q_{{\rm
T}0}\equiv Q_{\psi_{\rm T}\,0}$ and 
$Q_{0{\rm
T}}\equiv Q_{0\,\psi_{\rm T}}$. Using these relations, we can obtain
a matrix representation of $Q_{\Psi_0}\psi$:
\begin{eqnarray}
Q_{\Psi_0}\psi&=&\left(
 \begin{array}{cc}
P_0(Q_{\rm T}\chi)P_0 & P_0(Q_{{\rm T}0}\chi_b)\bar{\Sigma}_b\\
&\\
 \Sigma_a(Q_{0{\rm T}}\bar{\chi}_a)P_0 &
 \Sigma_a(Q_{\rm B}\phi_{ab})\bar{\Sigma}_b
\end{array}\right).
\end{eqnarray}
Consequently, 
the action expanded around $\Psi_0$ can be rewritten as 
\begin{eqnarray}
 S[\psi; Q_{\Psi_0}]&=& S[\phi_{ab};Q_{\rm B}]
+S'[\chi,\chi_a,\bar{\chi}_a,\phi_{ab}],
\label{phichiaction}
\end{eqnarray}
where each action is given by
\begin{eqnarray}
 S[\phi_{ab};Q_{\rm B}]&=&-\frac{1}{g^2}\int
\left(\frac{1}{2}\sum_{a=1}^N\sum_{b=1}^N
\phi_{ba}Q_{\rm B}\phi_{ab}
+\frac{1}{3}
\sum_{a=1}^N\sum_{b=1}^N\sum_{c=1}^N
\phi_{ab}\phi_{bc}\phi_{ca}\right)
\nn
&=&-\frac{1}{g^2}\int {\rm tr}
\left(\frac{1}{2}
\phi Q_{\rm B}\phi
+\frac{1}{3}
\phi^3\right),
\label{Sphi}
\end{eqnarray}
and
\begin{eqnarray}
S'[\chi,\chi_a,\bar{\chi}_a,\phi_{ab}]
&=&-\frac{1}{g^2}\int
\left(\frac{1}{2}\chi Q_{\rm T}\chi
+\sum_{a=1}^N\bar{\chi}_aQ_{{\rm T}0}\chi_a
+\frac{1}{3}\chi^3\right.
\nn
&&
\left.
+\sum_{a=1}^N\bar{\chi}_a\chi\chi_a
+\sum_{a=1}^N\sum_{b=1}^N\chi_a\phi_{ab}\bar{\chi}_b\right).
\label{Schi}
\end{eqnarray}
In (\ref{Sphi}), $\phi$ represents a matrix $(\phi_{ab})$ and the trace
denotes the sum of the diagonal elements with indices $a,b$.
Obviously, (\ref{Sphi}) represents the action for $N$ D-branes; namely,
$\phi_{ab}$ is a string field of an open string attached on
the $a$th and $b$th D-branes.
Moreover, in the action (\ref{Schi}),
$\chi$ is a string field on a D-brane with tachyon
condensation,
and $\chi_a$ and $\bar{\chi}_a$ represent string fields
of an open string attaching on a D-brane with tachyon condensation
and on one of the $N$ D-branes, on which $\phi_{ab}$ also attach.
Accordingly, the actions (\ref{Sphi}) and (\ref{Schi}) describe
the theory for $N+1$ D-branes in
which a D-brane vanishes due to tachyon condensation.
This system should be physically equivalent to the $N$ D-brane system 
because $Q_{\rm T}$ and $Q_{{\rm T}0}$ have trivial cohomology\footnote{
In Ref.~\cite{Ellwood:2006ba}, it is shown that 
a homotopy operator exists for $Q_{\rm T}$, $Q_{{\rm T}0}$, and $Q_{0{\rm
T}}$ if a homotopy state is given for $Q_{\rm T}$. 
For the identity-based tachyon vacuum solution\cite{Inatomi:2011xr}, 
$Q_{{\rm T}0}$
and $Q_{0{\rm T}}$ also have vanishing cohomology, as does $Q_{\rm T}$
\cite{Kishimoto:2002xi}, since a homotopy state can be constructed for
the solution.}
and therefore
 this result is consistent with the expectation that the solution
(\ref{EMsol}) is regarded as an $N$ D-brane solution.

Let us consider an on-shell closed string coupling to an open string
field. In the complex plane, a closed string vertex operator is given 
by ${\cal V}(z,\bar{z})=c(z)c(\bar{z})V_{\rm matt}(z,\bar{z})$, where
$V_{\rm matt}$ is a vertex operator with the conformal dimension $(1,1)$
in the matter sector.
We can give a BRST invariant state using ${\cal V}$ as
\begin{eqnarray}
 V={\cal V}(i,-i)I,
\end{eqnarray}
where the point $z=i$ corresponds to the midpoint of an open string.
Since the vertex is inserted at the midpoint, 
the state $V$ commutes with any string field $A$: $VA=AV$.
For the open string field $\Psi$,
an interaction term with the closed string vertex
is given as a gauge invariant overlap\cite{Zwiebach:1992bw}:
\begin{eqnarray}
 O_{\cal V}(\Psi)=\int V\Psi.
\end{eqnarray}
In the background of the $N$ D-brane solution, 
using (\ref{SigmaBarSigma}) and (\ref{PaSigma}),
we can easily find
couplings of the fluctuation fields to the closed string as
\begin{eqnarray}
 O_{\cal V}(\psi)
&=&O_{\cal V}(\chi)
+\sum_{a=1}^NO_{\cal V}(\phi_{aa}).
\end{eqnarray}
This correctly provides a closed string interaction to open strings
on the $N+1$ D-branes.


Next, we consider the correspondence between gauge symmetries in the
original action (\ref{action})
and the expanded action (\ref{phichiaction}).
The original gauge transformation is given by
\begin{eqnarray}
 \delta_\Lambda \Psi = Q_{\rm B}\Lambda+\Psi\Lambda-\Lambda\Psi.
\label{gaugetrans}
\end{eqnarray}
Since $\Psi=\Psi_0+\psi$,
 the gauge transformation for $\psi$ is given by
\begin{eqnarray}
 \delta_\Lambda\psi &=&
 Q_{\rm \Psi_0}\Lambda+\psi\Lambda-\Lambda\psi,
\label{gaugetrans3}
\end{eqnarray}
where we note that $\Lambda$ is the same parameter as in
(\ref{gaugetrans}).
Here, we decompose $\Lambda$ into
$\tilde{\Lambda}_{\alpha\beta}=P_\alpha\Lambda P_\beta$ by 
the projectors. Then, changing variables as
\begin{eqnarray}
 \tilde{\Lambda}_{ab}=\Sigma_a\Lambda_{ab}\bar{\Sigma}_b,
\ \ \ 
 \tilde{\Lambda}_{0a}=\lambda_a\bar{\Sigma}_a,
\ \ \ 
 \tilde{\Lambda}_{a0}=\Sigma_a\bar{\lambda}_a,
\end{eqnarray}
and writing $\tilde{\Lambda}_{00}=\lambda$, we find that
\begin{eqnarray}
 \delta_\Lambda \psi &=& \sum_{a=1}^N\sum_{b=1}^N
\Sigma_a(\delta_\Lambda \phi_{ab})\bar{\Sigma}_b
+\sum_{a=1}^N
P_0(\delta_\Lambda \chi_{a})\bar{\Sigma}_a
+\sum_{a=1}^N
\Sigma_a(\delta_\Lambda\bar{\chi}_{a})P_0
+P_0(\delta_\Lambda\chi)P_0,
\end{eqnarray}
where the gauge transformations for the components are given as
\begin{eqnarray}
 \delta_{\Lambda} \phi_{ab} &=& Q_{\rm
  B}\Lambda_{ab}+\phi_{ac}\Lambda_{cb} 
-\Lambda_{ac}\phi_{cb}
+\bar{\chi}_a\lambda_b 
-\bar{\lambda}_a\chi_b,
\nn
\delta_{\Lambda} \chi_a &=& P_0(Q_{{\rm T}0}\lambda_a)+\chi\lambda_a
+\chi_b\Lambda_{ba}
-\lambda\chi_a-\lambda_{b}\phi_{ba},
\nn
\delta_{\Lambda} \bar{\chi}_a 
&=& (Q_{0{\rm T}}\bar{\lambda}_a)P_0 
+\bar{\chi}_a\lambda+\phi_{ab}\bar{\lambda}_b
-\bar{\lambda}_a\chi-\Lambda_{ab}\bar{\chi}_b,
\nn
\delta_\Lambda \chi 
&=& P_0(Q_{\rm T}\lambda)P_0+\chi\lambda
+\chi_a\bar{\lambda}_a
-\lambda\chi-\lambda_a\bar{\chi}_a.
\end{eqnarray}


\section{Concluding remarks
\label{sec3}}

We have shown that the theory expanded around the $N$ D-brane solution
given by Erler-Maccaferri describes an $N+1$ D-brane system with a
vanishing
D-brane due to the tachyon condensation. By projectors made of
regularized bcc operators, an open string field in the original theory is
divided into multi-string fields with matrix indices. Then, these
indices can be regarded as Chan-Paton factors in the $N$ D-brane
background. We have found that $N^2$ string fields on $N$ D-branes are
embedded in a string field as block matrices. Similarly, gauge
transformation parameters in the expanded theory are represented as
block elements of a gauge parameter string field in the original
theory. 

From the matrix representation (\ref{psimatrix}), the string fields
$\tilde{\phi}_{\alpha\beta}$ are mutually independent variables and then
the degrees of freedom of $\tilde{\phi}_{\alpha\beta}$ are equivalent to
those of the string field $\psi$.
Then, it is natural to expect that the path integral measure of the
fluctuation $\psi$ is given by the product of measures of
$\tilde{\phi}_{\alpha\beta}$.
As seen in the previous section, 
we can rewrite $\tilde{\phi}_{\alpha\beta}$ as $\phi_{ab}$, 
$\chi$, $\chi_a$, and $\bar{\chi}_a$ by linear transformations.
Therefore, the measure of $\psi$ is expressed by the measures of the
string fields on the $N+1$ D-branes:
\begin{eqnarray}
  {\cal D}\psi
=\prod_{a=1}^N\prod_{b=1}^N {\cal
  D}\phi_{ab}\,{\cal D}\chi\,\prod_{a=1}^N{\cal D}\chi_a\,
\prod_{a=1}^N{\cal D}\bar{\chi}_a.
\label{measurephichi}
\end{eqnarray}
Hence, the matrix interpretation of open string fields ensures that
the quantum measure for the $N$ D-brane system is correctly
derived from the classical solution in the string field theory.

Finally, we should comment on the multiplicative ordering of $\Sigma_a$
and $\bar{\Sigma}_a$ in the projectors. As in (\ref{defP}), we have
defined the projectors such that $\Sigma_a$ does not operate on
$\bar{\Sigma}_a$, because bcc operators break associativity,
 as discussed in Ref.~\cite{Erler:2014eqa}.
To get a more definite result, we should separate these states by some
worldsheet. This is a similar approach to that adopted in
Ref.~\cite{Ishibashi:2014mua} to remedy the problem due
to another nonassociativity. Accordingly, we 
need to regularize $P_a$ by inserting some worldsheet between $\Sigma_a$
and $\bar{\Sigma}_a$. In the case that $\psi_{\rm T}$ is given by the
Erler-Schnabl solution, one possible choice for regularization is
\begin{eqnarray}
 P_a=\Sigma_a\,Q_{\rm
  T}\left(\frac{B}{1+K}e^{-\epsilon K}\right)\bar{\Sigma}_a,
\end{eqnarray}
where $\epsilon$ is a positive infinitesimal parameter. It is noted that
$B/(1+K)$ is a homotopy operator for $Q_{\rm T}$ and this construction
is parallel to that of the regularized bcc operators from $\sigma$ and
$\bar{\sigma}$ \cite{Erler:2014eqa}. It can easily be seen that
$P_aP_b=\delta_{ab}P_a$ and $Q_{\rm T}P_a=0$.

In this regularization, the limit $\epsilon\rightarrow 0$ should be
taken after calculating the correlation functions related to trace (or
integration) of string fields. It should never be done in string fields;
e.g., the state $P_a A$ keeps the parameter $\epsilon$
until correlation functions are calculated. Evidently, the state
with the regularization parameter is regarded as a kind of distribution
as in Ref.~\cite{Bonora:2013cya}
and indeed it is outside the usual Fock space like the
phantom term in Schnabl's tachyon vacuum solution \cite{Schnabl:2005gv}. 
We hope that, in terms of the projectors, it will be
possible to obtain a deeper understanding of a space of string fields,
in particular, the topology in the space beyond the single Fock space in
string field theories \cite{Bonora:2014mta}.

\section*{Acknowledgements}

The work of I. K. and T. T. is supported by a JSPS Grant-in-Aid for
Scientific Research (B) (\#24340051). The work of I. K. is supported in
part by a JSPS Grant-in-Aid for Young Scientists (B) (\#25800134).


\end{document}